\documentclass[english,aps,prl,twocolumn,amsmath,superscriptaddress,amssymb]{revtex4}

\usepackage[T1]{fontenc}
\usepackage[latin9]{inputenc}
\usepackage{amsthm}
\usepackage{amsmath}
\usepackage{color}
\makeatletter
\usepackage{graphicx}
\usepackage{dcolumn}
\usepackage{bm}
\usepackage{epsfig}
\usepackage[caption=false]{subfig} 

\makeatother

\usepackage{babel}

\newcommand{\diff}{\mathop{}\!\mathrm{d}}

\begin{document}

\title{Statistics of Extreme Waves in Random Media}

\author{Jakob J. Metzger}
\affiliation{Max Planck Institute for Dynamics and Self-Organization (MPIDS), Am Fassberg
17, 37077 G\"{o}ttingen, Germany}
\affiliation{Institute for Nonlinear Dynamics, Department of Physics, University
of Göttingen, Germany}
\author{Ragnar Fleischmann}
\affiliation{Max Planck Institute for Dynamics and Self-Organization (MPIDS), Am Fassberg
17, 37077 G\"{o}ttingen, Germany}
\author{Theo Geisel}
\affiliation{Max Planck Institute for Dynamics and Self-Organization (MPIDS), Am Fassberg
17, 37077 G\"{o}ttingen, Germany}
\affiliation{Institute for Nonlinear Dynamics, Department of Physics, University
of Göttingen, Germany}

\begin{abstract}
Waves traveling through random media exhibit random focusing that leads to extremely high wave intensities even in the absence of nonlinearities. Although such extreme events are present in a wide variety of physical systems and the statistics of the highest waves is important for their analysis and forecast, it remains poorly understood in particular in the regime where the waves are highest. We suggest a new approach that greatly simplifies the mathematical analysis and calculate the scaling and the distribution of the highest waves valid for a wide range of parameters. 
\end{abstract}

\maketitle

Extremely high waves can occur in virtually all systems in which waves travel through a complex medium. This holds in particular for wave propagation in media that can be considered as random, and includes phenomena such as the formation of freak or rogue waves in the ocean and the focusing of tsunamis  \cite{Ward:2002um,Berry:2005hi,Synolakis:2006ui,berry_focused_2007,heller_refraction_2008}, the branching of electron flows in semiconductor devices \cite{topinka_coherent_2001,jura_unexpected_2007,aidala_imaging_2007,maryenko_how_2012},  extreme sound waves in the ocean \cite{colosi_review_1999,wolfson_study_2000,wolfson_stability_2001}, rogue light waves \cite{arecchi_granularity_2011,akhmediev_editorial_2010,bonatto_deterministic_2011}, as well as the formation of so-called hot spots and branching in microwaves \cite{hohmann_freak_2010,Barkhofen:2013ta}.
Despite tremendous theoretical and experimental efforts over the last decades (see e.g. the reviews~\cite{barabanenkov_status_1971,kravtsov_propagation_1992,Kravtsov:1993tf,Rytov:1989kx,Klyatskin:2005uq}), however, the distribution of the highest waves remains elusive and the precise mechanism for their creation under debate. 

The formation of extreme waves has attracted attention from a broad range of researchers over many decades. A main difficulty in characterizing the distribution of these extreme events is due to the multiple processes involved in their formation. In the simplest model pioneered by Longuet-Higgins, extreme waves arise as accidentally coincident coherent superpositions of plane waves, whose probability (due to the central limit theorem) follows the tail of a Gaussian distribution \cite{LonguetHiggins:1957ut}. The latter disagrees, however, with the much higher probability with which extreme wave amplitudes have been observed empirically, e.g.~in all of the systems mentioned above. A further ingredient in many wave systems, e.g.~in water waves, are nonlinearities in the wave equations, which can locally increase the wave height. However, extreme waves can also be observed in systems that are very well described by \emph{linear} wave equations, leading to the conjecture that nonlinear effects merely further increase the effect of a process already present in linear wave equations. The most general mechanism generating heavy tails in the intensity distribution of linear waves is focusing by the collective effect of a correlated random medium leading to a branching of the wave flow~\cite{topinka_coherent_2001}. It occurs ubiquitously, as it is caused even by very weak disorder, which cannot be avoided in almost all real systems and originates, e.g., from donors in semiconductor nanostructures or from the intersection of currents in the ocean. 
The random focusing of the waves is the effect of the corresponding ray (or classical) dynamics and is known to lead to extremely high ray densities at focal points or lines, the  \emph{caustics} \cite{Berry:1980vk,kaplan_statistics_2002,Metzger:2013iq}.  As it is known that the extreme waves generated in this way are much more frequent than expected from the above Gaussian tails, it is crucial to mathematically quantify and predict their incidence. The complex dynamics of the rays, e.g.~the interplay of chaotic stretching and folding at the wavefront with the geometric singularities of the caustics and the intricate nature of diffraction and interference of the correlated random wave patterns, however, have so far defied attempts to derive the full distribution of the wave intensities in branched flows, and to establish the exact character of its heavy tail. To understand and describe the nature of linear rogue waves, we therefore take a new approach concentrating on the most pressing question: How high are the highest waves? More precisely, we want to study the highest waves in a cross-section of arbitrary but fixed length perpendicular to the main flow direction of the waves. 

In this Letter, we present an approach that maps the complex dynamics of the waves in the random medium to the simpler problem of an initially distorted wavefront that then travels through free space. By comparing relevant quantities in these two cases, we can derive a distribution for the extreme waves in systems governed by linear wave dynamics which is universal for a large class of random media, thus solving an important fundamental problem in understanding and predicting such events in a wide range of physical systems.

\begin{figure}[htbp]
	\centering
		\includegraphics[width=1\columnwidth]{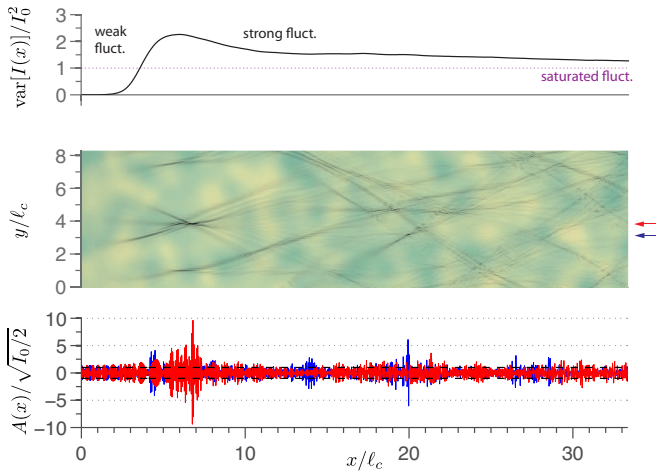}
	\caption{The formation of extreme waves in a random medium. (center panel): Illustration of a wave flow (gray scale) through a weak random medium (color), starting from an initially plane wave. Darker shades of gray correspond to higher flow intensity. The standard deviation of the random potential is $10\%$ of the energy of the flow, and the wavelength is $5\%$ of the correlation length of the potential, $\ell_c$. The image shows an extract of a larger simulation with periodic boundary conditions on top and bottom, and absorbing boundary conditions on the left and right. (top panel): The variance of the intensities averaged over 100 realizations of the random medium as a function of the distance from the source. A high variance of the wave intensity corresponds to the occurrence of extreme events. (bottom panel): Two sections of the wave amplitudes at the locations indicated by arrows in the central panel. The amplitude is normalized to the standard deviation of the fluctuations in the saturated regime far away from the source (thick black dashed lines). Wave amplitudes of up to $10$ times of this standard deviation can be observed. The probability to observe such an event in random waves according to Rayleigh's law would be smaller than $10^{-20}$.}
	\label{fig:fig1}
\end{figure}

The central panel of Fig.~\ref{fig:fig1} illustrates the formation of a branched flow and the associated occurrence of extreme waves. We study linear wave propagation by using the paradigmatic case of quantum mechanical waves described by the time-independent Schr\"odinger equation propagating in a two-dimensional weak random medium modeled as a Gaussian random field. Starting from a monochromatic unidirectional plane wave source, strong intensity fluctuations appear at a characteristic length-scale well below the mean free path. Generalizations of this initial condition have been studied e.g.~in \cite{heller_refraction_2008,Ying:2011fp}. The regime studied here is usually classified as the regime of \emph{strong fluctuations} which emphasizes the high probability of extreme events occurring. It is most easily characterized by the variance of the squared wave heights or intensities \cite{Barkhofen:2013ta}, which is much higher than in the weak fluctuation regime close to the wave source or the regime of saturated intensity fluctuations, where the distribution of the intensities is exponential (Rayleigh's law) \cite{LonguetHiggins:1957ut,Metzger:2013iq} (see upper panel of Fig.~\ref{fig:fig1}).

In all of the applications mentioned above, the random medium is typically modeled as a Gaussian random field $V(\mathbf{r})$, where $\mathbf{r}$ is a position vector, and whose properties are defined by its isotropic spatial correlation function $c(\mathbf{r}) = \langle V(\mathbf{r'}+\mathbf{r}) V(\mathbf{r'})\rangle$. The medium is assumed to have a spatial correlation length $\ell_c$ and a dimensionless amplitude relative to the kinetic energy of the flow, $\epsilon$. The crucial wave parameter is the wavelength $\lambda$, which we assume to be smaller than the typical scale of the random medium $\ell_c$. This is a prerequisite for the waves to be able to resolve the structure of the random medium and therefore for focusing to occur. 

The central question of this work is to understand whether there is a universal mathematical description for the distribution of extreme waves in different random media and how the influence of the medium parameters $\epsilon$ and $\ell_c$ as well as the wavelength $\lambda$ is reflected in the distribution. Our main conjecture is that the extreme waves occur at locations where the waves are focused, which in two dimensions happens generically at a \emph{fold line}, which is known from e.g.~the aberration of an imperfect lens in optics. It should then suffice to understand the wave dynamics and its dependence on the system parameters in the close vicinity of the focal lines, where the wave structure has a generic interference and diffraction pattern that can be understood using \emph{diffraction integrals} \cite{Berry:1980vk,nye1999natural}. 

The diffraction integral yielding the wave pattern perpendicular to a fold line has been derived in the framework of catastrophe optics \cite{Berry:1980vk,nye1999natural},
and is given by

\begin{equation}
   J(\xi) = \int_{-\infty}^{\infty} \diff y_0 \, e^{-i\,k\,\phi(\xi,y_0)}
   \label{eqn:di}
\end{equation}
where $k=2\pi/\lambda$ is the wave number, $\xi = y/R$ a dimensionless spatial coordinate perpendicular to the fold line rescaled by the distance $R$ from the source, and $\phi(\xi,y_0) = \frac{1}{3}\alpha\,y_0^3 + \xi\,y_0$ is the normal form of the fold caustic, which can be interpreted as the optical path length from the source to the observation point. The parameter $\alpha$ with units of inverse length squared describes the shape of the fold and will be treated in more detail below. Expression \eqref{eqn:di} can be derived by assuming a uniform medium and a wave propagating in $x$-direction with an initial waveform slightly deformed from the plane wave, which, because of its initial curvature, focuses at a certain distance from the source. A suitably curved initial wavefront that produces the normal form of the optical path $\phi(\xi,y_0)$ given above is described by its deviation $f(y_0)$ from a straight line at $t=0$ (such that $x_0=f(y_0)$ at $t=0$) and is given by $f(y_0) = \frac{y_0^{2}}{2R}+\frac{1}{3}\alpha y_0^{3}$.
Assuming the normal form and initial condition given above, it can be shown that the expression for the wave function across a fold is given by \cite{nye1999natural}
\begin{equation}
	\psi(\xi) =\left(\frac{2\pi  k}{i R}\right)^{1/2}\left(k\alpha\right)^{-1/3}e^{i k R} \mathrm{Ai}\left(k^{2/3} \alpha^{-1/3} \xi\right),
	\label{eqn:di2}
\end{equation}
where $\mathrm{Ai}$ denotes the Airy function of the first kind. We will use the prefactor of this dimensionless wave function to calculate the scaling of the highest waves without having to integrate over a spatial coordinate.

\begin{figure}[htbp]
	\centering
		\includegraphics[width=\columnwidth]{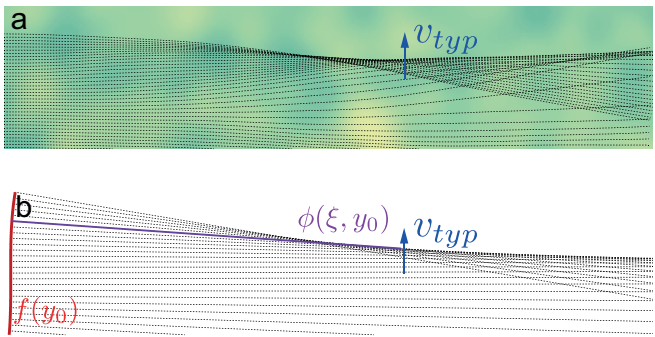}
	\caption{Mapping the complex dynamics of the propagation in the random medium to the simpler problem of an initially curved wavefront propagating through free space. (a) An initially plane ray bundle propagating through a random medium develops a fold caustic, with typical transverse velocity $v_{typ}$. (b) In a situation mimicking the random ray propagation, a ray bundle emanating perpendicularly to an initially curved wavefront described by a deviation $f(y_0)$ from the plane initial condition along the initial $y_0$-axis propagates through free space (see text for details). The path length corresponding to $f(y_0)$ is given by $\phi(\xi,y_0)$. By matching the velocities $v_{typ}$ of the ray bundles in both cases the local caustic parameter $\alpha$ can be related to the parameters of the stochastic medium.}
	\label{fig:fig2a}
\end{figure}

The normal form $\phi(\xi,y_0)$  is an expression for the characteristic variation of  the \emph{optical path length} of the rays that form the fold caustic. In our case, however, this variation of the optical path length is due to the fluctuations of the medium and not due to the shape of the initial wavefront. The next step is therefore to develop a way of relating the local parameter $\alpha$ to the global properties of the random wave flow, and thus to the stochastic properties of the medium. This is illustrated in Fig.~\ref{fig:fig2a} and can be carried out by examining the velocity profile of the ray bundle generated by $\phi$.  First, we observe that in the random medium the distance $R$ to the focal region is given by the average propagation distance to a focal line or caustic, i.e.~$R=x_c$. Furthermore, since the rays are orthogonal to the wavefront, the velocity in the transverse $y$-direction is given by $v_y = -\partial_y f(y) = - y/x_c-\alpha y^{2}$. We set the initial velocity in the longitudinal direction, $v_0$, to unity and express all velocities in units of this, thus making $v_y$ dimensionless. At the caustics (which are turning points), the velocity has an extremum, which is obtained at $y^*=-(2\alpha x_c)^{-1}$. Plugging this into the expression for the velocity we obtain a typical velocity of the rays at the caustics of $v_{typ} = v_y(y^*) = (4 \alpha x_c^2)^{-1}$. Thus the parameter $\alpha$ describing the shape of the caustic is related to the properties of the random medium through 
\begin{equation}
	\alpha = (4 v_{typ} x_c^2)^{-1}.
	\label{eq:alpha}
\end{equation}

The typical velocity $v_{typ}$ and the typical distance to a caustic $x_c$ at which this velocity should be evaluated can be obtained from a stochastic model of the ray propagation associated with the waves as follows.

Assuming that the random medium is weak, the ray equations can be reduced to one-dimensional equations describing the dynamics in the transverse direction, with the propagation in the longitudinal direction now playing the role of time (paraxial approximation).
Also, the random potential $V(\mathbf{r})$ can be approximated as uncorrelated Gaussian white noise of appropriate strength $\sigma$, such that the ray equations become (with the particle mass set to unity) $\dot{y}/v_0 = v_y$ and $\dot{v}_y = \sigma\, \zeta(t)$ with $\langle \zeta(t)\zeta(t') \rangle=\delta(t-t')$. The prefactor of the noise term, $\sigma$, now encodes the properties of the random medium and can be calculated from its correlation function $c(\mathbf{r}) = \langle V(\mathbf{r'}+\mathbf{r}) V(\mathbf{r'})\rangle$ as $\sigma^2 = v_0^{-3} \int_{-\infty}^{\infty} \diff x \,\partial_{yy}c(x,y) \big|_{y=0}$ \cite{Kulkarny:1982tm,wolfson_stability_2001,metzger_universal_2010}, where the initial velocity $v_0$ is kept for dimensional clarity, but since it is equal to unity in the paraxial approximation, will be dropped from now on. 

Since all moments of the random processes $y$ and $v_y$ can easily be calculated, they can serve as estimates for the quantities $v_{typ}$ and $x_c$. The typical time to the first caustics can be calculated assuming that they occur when the fluctuations of the rays in the transverse direction are on the order of the correlation length of the potential, i.e.~$\langle y^2(t_c)\rangle \sim \sigma^2 t_c^3 /3 \sim \ell_c^2$ which results in a scaling of $t_c \sim \epsilon^{-2/3}\ell_c$ \cite{kaplan_statistics_2002} in accordance with more detailed calculations \cite{Kulkarny:1982tm, metzger_universal_2010}. This is the typical time and in the paraxial approximation also the distance the rays travel until the first focal points are reached. We note that in weak random media this happens on a shorter length scale than the mean free path, which scales as $\epsilon^{-2}\ell_c$. As the simplest estimate for the typical transverse velocity of the rays in the region where the extreme waves are expected, we assume a scaling as $v_{typ} = \sqrt{\langle v^2(t_c)\rangle} \sim \sqrt{ 2\sigma^2 t_c } \sim \epsilon^{2/3}$. 

Using Eqs.~\eqref{eqn:di2} and \eqref{eq:alpha}, the scaling of the wave intensities at the fold, $I_{\mathrm{fold}}$, takes the form
\begin{equation}
	I_{\mathrm{fold}}=\left|\psi_{\mathrm{fold}}^2(\lambda,\epsilon,\ell_c)\right| = c\,(\lambda/\ell_c)^{-1/3} \epsilon^{2/9}
	\label{eq:i}
\end{equation}
where $c$ is a universal constant that can later be absorbed in the full distribution. We note that the fact that the scaling depends on the \emph{relative} wavelength $\lambda/\ell_c$ is quite intuitive considering that $\ell_c$ defines the spatial scale of the problem. We also note that the interesting scaling of $2/9$ with the strength of the random potential was, although not stated explicitly and using a different approach, anticipated by Kaplan \cite{kaplan_statistics_2002} for the more restricted case of the intensity of the branches that can just be resolved for a given wavelength.
Having linked the global flow properties determined by the parameters of the random medium with the local structure of the focal regions, which results in the scaling laws of Eq.~\eqref{eq:i}, we can now turn to the actual distribution of the extreme waves. 

At an arbitrary but, in proportion to $x_c$, fixed distance from the source within the strong fluctuation regime, we divide the flow intensities along a line transverse to the flow into sections of a few correlation lengths and, for each section, obtain the maximum intensity $I_{max}$. The precise number of correlation lengths is not important, the interval only has to be long enough so that it typically covers at least one branch, i.e.~one focal region. For the simulation used here we assume a Gaussian correlation function for the random potential, expecting, however, our results to also hold for different correlation functions because of the universality of the branch statistics \cite{metzger_universal_2010}.
Since for Gaussian correlated potentials there is on average approximately one focal region per six correlation lengths \cite{metzger_universal_2010}, we here take the maximum over ten correlation lengths to ensure that in most sections there is an extreme wave. We note that although there are not a large number of caustics across which each maximum is taken, there is, in each bin, a large number of independent intensity fluctuations which are the result of the focusing and interference of different parts of the wavefront which have traveled across different regions of the random medium. Therefore, the maximum is taken over many independent realizations of intensities and can be expected to follow an extreme value distribution, as will be confirmed by our numerical results below. Additionally, we have confirmed that our results hold if a different number of correlation lengths is used and also that it holds for different distances from the source in the strong fluctuation regime.

\begin{figure}[htbp]
	\centering
		\includegraphics[width=.8\columnwidth]{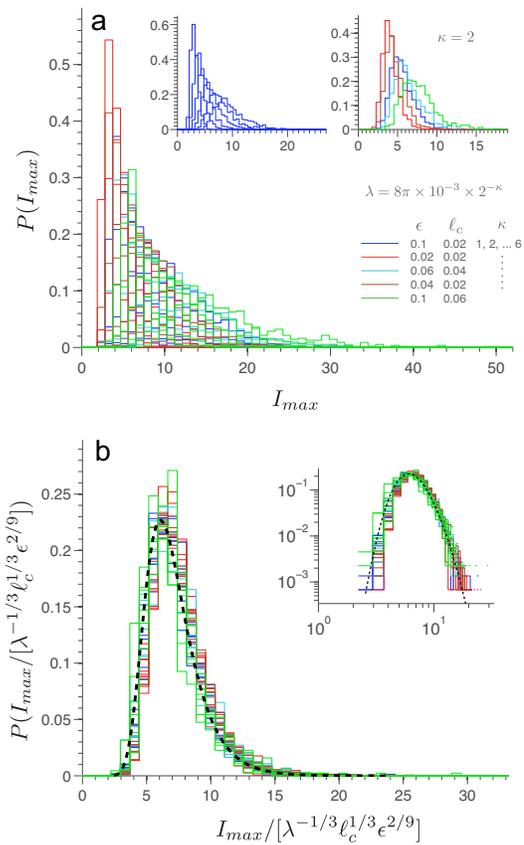}
	\caption{Distribution of extreme waves. (a) Unscaled distributions for different wavelengths $\lambda$ and different parameters of the random medium, $\epsilon$ (standard deviation) and $\ell_c$ (correlation length). The left inset displays curves for fixed medium parameters but varying wavelengths, while the right inset shows curves for varying medium parameters at a constant wavelength. (b) The same curves, but rescaled according to Eq.~\eqref{eq:fulldistr}, together with the distribution,  Eq.~\eqref{eq:fulldistr} (dashed line). The inset shows the same curves in a double-logarithmic plot.}
	\label{fig:fig2}
\end{figure}

Since the focal points cause a heavy-tailed intensity profile, the appropriate extreme value distribution is the Fr\'{e}chet distribution \cite{DeHaan:2006bi,Albeverio:2006up}. Thus, the full cumulative probability distribution of the maximum wave heights is given by
\begin{equation}
	P(I/[ \lambda^{-1/3}\ell_c^{1/3}\epsilon^{2/9}] \leq I_{max}) = e^{-[(I_{max}-m)/s]^{-a}}
	\label{eq:fulldistr}
\end{equation}
where the location, scale and shape parameters $m$, $s$ and $a$ are numerical constants independent of the random medium and only have to be measured once numerically. We confirm our results with detailed numerical simulations of wave flows and media with a wide range of parameters $\lambda$, $\epsilon$ and $\ell_c$. Fig.~\ref{fig:fig2} shows the numerical results of the distribution of the heights of the extreme waves for different sets of parameters. The datasets are histograms of the maximum wave heights across ten correlation lengths obtained from 200 realizations of the random potential (each extending on the order of a hundred correlation lengths) for each parameter set. The histograms are plotted together with an appropriate Fr\'{e}chet distribution with parameters $a=62.5$, $m=-95.2$, $s=101.3$. We note that the Fr\'{e}chet distribution describes the data very well, validating our assumptions of the independence of the individual maxima as well as the assumption that they are drawn from an underlying heavy-tailed distribution. Most importantly, we furthermore observe a striking data collapse when the data is scaled according to Eq.~\eqref{eq:i}  confirming the full universal distribution derived in Eq.~\eqref{eq:fulldistr}. As can be expected from the derivation above, the prediction is particularly accurate for the highest waves, i.e. in the tail of the distribution. Interestingly, we have observed the parameter values of the Fr\'{e}chet distribution to be remarkably constant in the whole strong fluctuation region indicated in Fig.~\ref{fig:fig1}, even though eventually the parameters will have to change with propagation distance, as the regime of exponentially distributed intensities is approached.

In conclusion, we have obtained the scaling behavior and the distribution of extreme waves in a random medium by analyzing the local wave profile at focal regions and by linking this to stochastic properties of the flow. We expect our results to be applicable in the analysis and prediction of extreme waves in a wide range of physical systems. Moreover, the results serve as a baseline for disentangling and estimating the effect of nonlinearities on extreme wave heights, and an extension of the method could directly include nonlinear effects. The methodology presented here also presents a viable route for analyzing further properties of branched flows and, more generally, wave propagation in random media.

\acknowledgments{This work has been supported by the DFG research group 760.}


\begin{thebibliography}{32}
\expandafter\ifx\csname natexlab\endcsname\relax\def\natexlab#1{#1}\fi
\expandafter\ifx\csname bibnamefont\endcsname\relax
  \def\bibnamefont#1{#1}\fi
\expandafter\ifx\csname bibfnamefont\endcsname\relax
  \def\bibfnamefont#1{#1}\fi
\expandafter\ifx\csname citenamefont\endcsname\relax
  \def\citenamefont#1{#1}\fi
\expandafter\ifx\csname url\endcsname\relax
  \def\url#1{\texttt{#1}}\fi
\expandafter\ifx\csname urlprefix\endcsname\relax\def\urlprefix{URL }\fi
\providecommand{\bibinfo}[2]{#2}
\providecommand{\eprint}[2][]{\url{#2}}

\bibitem[{\citenamefont{Ward}(2002)}]{Ward:2002um}
\bibinfo{author}{\bibfnamefont{S.~N.} \bibnamefont{Ward}}
  (\bibinfo{publisher}{Elsevier}, \bibinfo{address}{Amsterdam},
  \bibinfo{year}{2002}), pp. \bibinfo{pages}{175--191}.

\bibitem[{\citenamefont{Berry}(2005)}]{Berry:2005hi}
\bibinfo{author}{\bibfnamefont{M.~V.} \bibnamefont{Berry}},
  \bibinfo{journal}{New Journal of Physics} \textbf{\bibinfo{volume}{7}},
  \bibinfo{pages}{129} (\bibinfo{year}{2005}).

\bibitem[{\citenamefont{Synolakis and Bernard}(2006)}]{Synolakis:2006ui}
\bibinfo{author}{\bibfnamefont{C.~E.} \bibnamefont{Synolakis}}
  \bibnamefont{and} \bibinfo{author}{\bibfnamefont{E.~N.}
  \bibnamefont{Bernard}}, \bibinfo{journal}{Philosophical Transactions of the
  Royal Society A: Mathematical, Physical and Engineering Sciences}
  \textbf{\bibinfo{volume}{364}}, \bibinfo{pages}{2231} (\bibinfo{year}{2006}).

\bibitem[{\citenamefont{Berry}(2007)}]{berry_focused_2007}
\bibinfo{author}{\bibfnamefont{M.~V.} \bibnamefont{Berry}},
  \bibinfo{journal}{Proceedings of the Royal Society A: Mathematical, Physical
  and Engineering Sciences} \textbf{\bibinfo{volume}{463}},
  \bibinfo{pages}{3055} (\bibinfo{year}{2007}).

\bibitem[{\citenamefont{Heller et~al.}(2008)\citenamefont{Heller, Kaplan, and
  Dahlen}}]{heller_refraction_2008}
\bibinfo{author}{\bibfnamefont{E.~J.} \bibnamefont{Heller}},
  \bibinfo{author}{\bibfnamefont{L.}~\bibnamefont{Kaplan}}, \bibnamefont{and}
  \bibinfo{author}{\bibfnamefont{A.}~\bibnamefont{Dahlen}},
  \bibinfo{journal}{Journal of Geophysical Research}
  \textbf{\bibinfo{volume}{113}}, \bibinfo{pages}{C09023}
  (\bibinfo{year}{2008}).

\bibitem[{\citenamefont{Topinka et~al.}(2001)\citenamefont{Topinka, LeRoy,
  Westervelt, Shaw, Fleischmann, Heller, Maranowski, and
  Gossard}}]{topinka_coherent_2001}
\bibinfo{author}{\bibfnamefont{M.~A.} \bibnamefont{Topinka}},
  \bibinfo{author}{\bibfnamefont{B.~J.} \bibnamefont{LeRoy}},
  \bibinfo{author}{\bibfnamefont{R.~M.} \bibnamefont{Westervelt}},
  \bibinfo{author}{\bibfnamefont{S.~E.~J.} \bibnamefont{Shaw}},
  \bibinfo{author}{\bibfnamefont{R.}~\bibnamefont{Fleischmann}},
  \bibinfo{author}{\bibfnamefont{E.~J.} \bibnamefont{Heller}},
  \bibinfo{author}{\bibfnamefont{K.~D.} \bibnamefont{Maranowski}},
  \bibnamefont{and} \bibinfo{author}{\bibfnamefont{A.~C.}
  \bibnamefont{Gossard}}, \bibinfo{journal}{Nature}
  \textbf{\bibinfo{volume}{410}}, \bibinfo{pages}{183} (\bibinfo{year}{2001}).

\bibitem[{\citenamefont{Jura et~al.}(2007)\citenamefont{Jura, Topinka, Urban,
  Yazdani, Shtrikman, Pfeiffer, West, and
  Goldhaber-Gordon}}]{jura_unexpected_2007}
\bibinfo{author}{\bibfnamefont{M.~P.} \bibnamefont{Jura}},
  \bibinfo{author}{\bibfnamefont{M.~A.} \bibnamefont{Topinka}},
  \bibinfo{author}{\bibfnamefont{L.}~\bibnamefont{Urban}},
  \bibinfo{author}{\bibfnamefont{A.}~\bibnamefont{Yazdani}},
  \bibinfo{author}{\bibfnamefont{H.}~\bibnamefont{Shtrikman}},
  \bibinfo{author}{\bibfnamefont{L.~N.} \bibnamefont{Pfeiffer}},
  \bibinfo{author}{\bibfnamefont{K.~W.} \bibnamefont{West}}, \bibnamefont{and}
  \bibinfo{author}{\bibfnamefont{D.}~\bibnamefont{Goldhaber-Gordon}},
  \bibinfo{journal}{Nature Physics} \textbf{\bibinfo{volume}{3}},
  \bibinfo{pages}{841} (\bibinfo{year}{2007}).

\bibitem[{\citenamefont{Aidala et~al.}(2007)\citenamefont{Aidala, Parrott,
  Kramer, Heller, Westervelt, Hanson, and Gossard}}]{aidala_imaging_2007}
\bibinfo{author}{\bibfnamefont{K.~E.} \bibnamefont{Aidala}},
  \bibinfo{author}{\bibfnamefont{R.~E.} \bibnamefont{Parrott}},
  \bibinfo{author}{\bibfnamefont{T.}~\bibnamefont{Kramer}},
  \bibinfo{author}{\bibfnamefont{E.~J.} \bibnamefont{Heller}},
  \bibinfo{author}{\bibfnamefont{R.~M.} \bibnamefont{Westervelt}},
  \bibinfo{author}{\bibfnamefont{M.~P.} \bibnamefont{Hanson}},
  \bibnamefont{and} \bibinfo{author}{\bibfnamefont{A.~C.}
  \bibnamefont{Gossard}}, \bibinfo{journal}{Nature Physics}
  \textbf{\bibinfo{volume}{3}}, \bibinfo{pages}{464} (\bibinfo{year}{2007}).

\bibitem[{\citenamefont{Maryenko et~al.}(2012)\citenamefont{Maryenko, Ospald,
  von Klitzing, Smet, Metzger, Fleischmann, Geisel, and
  Umansky}}]{maryenko_how_2012}
\bibinfo{author}{\bibfnamefont{D.}~\bibnamefont{Maryenko}},
  \bibinfo{author}{\bibfnamefont{F.}~\bibnamefont{Ospald}},
  \bibinfo{author}{\bibfnamefont{K.}~\bibnamefont{von Klitzing}},
  \bibinfo{author}{\bibfnamefont{J.}~\bibnamefont{Smet}},
  \bibinfo{author}{\bibfnamefont{J.~J.} \bibnamefont{Metzger}},
  \bibinfo{author}{\bibfnamefont{R.}~\bibnamefont{Fleischmann}},
  \bibinfo{author}{\bibfnamefont{T.}~\bibnamefont{Geisel}}, \bibnamefont{and}
  \bibinfo{author}{\bibfnamefont{V.}~\bibnamefont{Umansky}},
  \bibinfo{journal}{Physical Review B} \textbf{\bibinfo{volume}{85}},
  \bibinfo{pages}{195329} (\bibinfo{year}{2012}).

\bibitem[{\citenamefont{Colosi et~al.}(1999)\citenamefont{Colosi, Baggeroer,
  Birdsall, Clark, Cornuelle, Costa, Dushaw, Dzieciuch, Forbes, Howe
  et~al.}}]{colosi_review_1999}
\bibinfo{author}{\bibfnamefont{J.~A.} \bibnamefont{Colosi}},
  \bibinfo{author}{\bibfnamefont{A.~B.} \bibnamefont{Baggeroer}},
  \bibinfo{author}{\bibfnamefont{T.~G.} \bibnamefont{Birdsall}},
  \bibinfo{author}{\bibfnamefont{C.}~\bibnamefont{Clark}},
  \bibinfo{author}{\bibfnamefont{B.~D.} \bibnamefont{Cornuelle}},
  \bibinfo{author}{\bibfnamefont{D.}~\bibnamefont{Costa}},
  \bibinfo{author}{\bibfnamefont{B.~D.} \bibnamefont{Dushaw}},
  \bibinfo{author}{\bibfnamefont{M.~A.} \bibnamefont{Dzieciuch}},
  \bibinfo{author}{\bibfnamefont{A.~M.~G.} \bibnamefont{Forbes}},
  \bibinfo{author}{\bibfnamefont{B.~M.} \bibnamefont{Howe}},
  \bibnamefont{et~al.}, \bibinfo{journal}{Oceanic Engineering, IEEE Journal of}
  \textbf{\bibinfo{volume}{24}}, \bibinfo{pages}{138{\textendash}155}
  (\bibinfo{year}{1999}).

\bibitem[{\citenamefont{Wolfson and Tappert}(2000)}]{wolfson_study_2000}
\bibinfo{author}{\bibfnamefont{M.~A.} \bibnamefont{Wolfson}} \bibnamefont{and}
  \bibinfo{author}{\bibfnamefont{F.~D.} \bibnamefont{Tappert}},
  \bibinfo{journal}{The Journal of the Acoustical Society of America}
  \textbf{\bibinfo{volume}{107}}, \bibinfo{pages}{154} (\bibinfo{year}{2000}).

\bibitem[{\citenamefont{Wolfson and Tomsovic}(2001)}]{wolfson_stability_2001}
\bibinfo{author}{\bibfnamefont{M.~A.} \bibnamefont{Wolfson}} \bibnamefont{and}
  \bibinfo{author}{\bibfnamefont{S.}~\bibnamefont{Tomsovic}},
  \bibinfo{journal}{The Journal of the Acoustical Society of America}
  \textbf{\bibinfo{volume}{109}}, \bibinfo{pages}{2693} (\bibinfo{year}{2001}).

\bibitem[{\citenamefont{Arecchi et~al.}(2011)\citenamefont{Arecchi, Bortolozzo,
  Montina, and Residori}}]{arecchi_granularity_2011}
\bibinfo{author}{\bibfnamefont{F.~T.} \bibnamefont{Arecchi}},
  \bibinfo{author}{\bibfnamefont{U.}~\bibnamefont{Bortolozzo}},
  \bibinfo{author}{\bibfnamefont{A.}~\bibnamefont{Montina}}, \bibnamefont{and}
  \bibinfo{author}{\bibfnamefont{S.}~\bibnamefont{Residori}},
  \bibinfo{journal}{Physical Review Letters} \textbf{\bibinfo{volume}{106}},
  \bibinfo{pages}{153901} (\bibinfo{year}{2011}).

\bibitem[{\citenamefont{Akhmediev and
  Pelinovsky}(2010)}]{akhmediev_editorial_2010}
\bibinfo{author}{\bibfnamefont{N.}~\bibnamefont{Akhmediev}} \bibnamefont{and}
  \bibinfo{author}{\bibfnamefont{E.}~\bibnamefont{Pelinovsky}},
  \bibinfo{journal}{The European Physical Journal Special Topics}
  \textbf{\bibinfo{volume}{185}}, \bibinfo{pages}{1} (\bibinfo{year}{2010}).

\bibitem[{\citenamefont{Bonatto et~al.}(2011)\citenamefont{Bonatto, Feyereisen,
  Barland, Giudici, Masoller, Leite, and
  Tredicce}}]{bonatto_deterministic_2011}
\bibinfo{author}{\bibfnamefont{C.}~\bibnamefont{Bonatto}},
  \bibinfo{author}{\bibfnamefont{M.}~\bibnamefont{Feyereisen}},
  \bibinfo{author}{\bibfnamefont{S.}~\bibnamefont{Barland}},
  \bibinfo{author}{\bibfnamefont{M.}~\bibnamefont{Giudici}},
  \bibinfo{author}{\bibfnamefont{C.}~\bibnamefont{Masoller}},
  \bibinfo{author}{\bibfnamefont{J.~R.~R.} \bibnamefont{Leite}},
  \bibnamefont{and} \bibinfo{author}{\bibfnamefont{J.~R.}
  \bibnamefont{Tredicce}}, \bibinfo{journal}{Physical Review Letters}
  \textbf{\bibinfo{volume}{107}}, \bibinfo{pages}{053901}
  (\bibinfo{year}{2011}).

\bibitem[{\citenamefont{H{\"o}hmann et~al.}(2010)\citenamefont{H{\"o}hmann,
  Kuhl, St{\"o}ckmann, Kaplan, and Heller}}]{hohmann_freak_2010}
\bibinfo{author}{\bibfnamefont{R.}~\bibnamefont{H{\"o}hmann}},
  \bibinfo{author}{\bibfnamefont{U.}~\bibnamefont{Kuhl}},
  \bibinfo{author}{\bibfnamefont{H.~J.} \bibnamefont{St{\"o}ckmann}},
  \bibinfo{author}{\bibfnamefont{L.}~\bibnamefont{Kaplan}}, \bibnamefont{and}
  \bibinfo{author}{\bibfnamefont{E.~J.} \bibnamefont{Heller}},
  \bibinfo{journal}{Physical Review Letters} \textbf{\bibinfo{volume}{104}},
  \bibinfo{pages}{093901} (\bibinfo{year}{2010}).

\bibitem[{\citenamefont{Barkhofen et~al.}(2013)\citenamefont{Barkhofen,
  Metzger, Fleischmann, Kuhl, and St{\"o}ckmann}}]{Barkhofen:2013ta}
\bibinfo{author}{\bibfnamefont{S.}~\bibnamefont{Barkhofen}},
  \bibinfo{author}{\bibfnamefont{J.~J.} \bibnamefont{Metzger}},
  \bibinfo{author}{\bibfnamefont{R.}~\bibnamefont{Fleischmann}},
  \bibinfo{author}{\bibfnamefont{U.}~\bibnamefont{Kuhl}}, \bibnamefont{and}
  \bibinfo{author}{\bibfnamefont{H.~J.} \bibnamefont{St{\"o}ckmann}},
  \bibinfo{journal}{Physical Review Letters} \textbf{\bibinfo{volume}{111}},
  \bibinfo{pages}{183902} (\bibinfo{year}{2013}).

\bibitem[{\citenamefont{Barabanenkov et~al.}(1971)\citenamefont{Barabanenkov,
  Kravtsov, Rytov, and Tamarski{\u \i}}}]{barabanenkov_status_1971}
\bibinfo{author}{\bibfnamefont{Y.~N.} \bibnamefont{Barabanenkov}},
  \bibinfo{author}{\bibfnamefont{Y.~A.} \bibnamefont{Kravtsov}},
  \bibinfo{author}{\bibfnamefont{S.~M.} \bibnamefont{Rytov}}, \bibnamefont{and}
  \bibinfo{author}{\bibfnamefont{V.~I.} \bibnamefont{Tamarski{\u \i}}},
  \bibinfo{journal}{Soviet Physics Uspekhi} \textbf{\bibinfo{volume}{13}},
  \bibinfo{pages}{551} (\bibinfo{year}{1971}).

\bibitem[{\citenamefont{Kravtsov}(1992)}]{kravtsov_propagation_1992}
\bibinfo{author}{\bibfnamefont{Y.~A.} \bibnamefont{Kravtsov}},
  \bibinfo{journal}{Reports on Progress in Physics}
  \textbf{\bibinfo{volume}{55}}, \bibinfo{pages}{39} (\bibinfo{year}{1992}).

\bibitem[{\citenamefont{Kravtsov}(1993)}]{Kravtsov:1993tf}
\bibinfo{author}{\bibfnamefont{Y.~A.} \bibnamefont{Kravtsov}},
  \bibinfo{journal}{Applied Optics} \textbf{\bibinfo{volume}{32}},
  \bibinfo{pages}{2681} (\bibinfo{year}{1993}).

\bibitem[{\citenamefont{Rytov et~al.}(1989)\citenamefont{Rytov, Kravtsov, and
  Tatarskii}}]{Rytov:1989kx}
\bibinfo{author}{\bibfnamefont{S.~M.} \bibnamefont{Rytov}},
  \bibinfo{author}{\bibfnamefont{Y.~A.} \bibnamefont{Kravtsov}},
  \bibnamefont{and} \bibinfo{author}{\bibfnamefont{V.~I.}
  \bibnamefont{Tatarskii}}, \emph{\bibinfo{title}{{Principles of Statistical
  Radiophysics: Wave propagation through random media }}},
  vol.~\bibinfo{volume}{4} (\bibinfo{publisher}{Springer},
  \bibinfo{address}{Berlin}, \bibinfo{year}{1989}).

\bibitem[{\citenamefont{Klyatskin}(2005)}]{Klyatskin:2005uq}
\bibinfo{author}{\bibfnamefont{V.~I.} \bibnamefont{Klyatskin}},
  \emph{\bibinfo{title}{{Stochastic Equations Through the Eye of the Physicist:
  Basic Concepts, Exact Results And Asymptotic Approximations}}}
  (\bibinfo{publisher}{Elsevier Science Limited}, \bibinfo{year}{2005}).

\bibitem[{\citenamefont{Longuet-Higgins}(1957)}]{LonguetHiggins:1957ut}
\bibinfo{author}{\bibfnamefont{M.~S.} \bibnamefont{Longuet-Higgins}},
  \bibinfo{journal}{Philosophical Transactions of the Royal Society of London.
  Series A, Mathematical and Physical Sciences} \textbf{\bibinfo{volume}{249}},
  \bibinfo{pages}{321} (\bibinfo{year}{1957}).

\bibitem[{\citenamefont{Berry and Upstill}(1980)}]{Berry:1980vk}
\bibinfo{author}{\bibfnamefont{M.~V.} \bibnamefont{Berry}} \bibnamefont{and}
  \bibinfo{author}{\bibfnamefont{C.}~\bibnamefont{Upstill}},
  \bibinfo{journal}{Prog. Opt} \textbf{\bibinfo{volume}{18}},
  \bibinfo{pages}{257} (\bibinfo{year}{1980}).

\bibitem[{\citenamefont{Kaplan}(2002)}]{kaplan_statistics_2002}
\bibinfo{author}{\bibfnamefont{L.}~\bibnamefont{Kaplan}},
  \bibinfo{journal}{Physical Review Letters} \textbf{\bibinfo{volume}{89}},
  \bibinfo{pages}{184103} (\bibinfo{year}{2002}).

\bibitem[{\citenamefont{Metzger et~al.}(2013)\citenamefont{Metzger,
  Fleischmann, and Geisel}}]{Metzger:2013iq}
\bibinfo{author}{\bibfnamefont{J.~J.} \bibnamefont{Metzger}},
  \bibinfo{author}{\bibfnamefont{R.}~\bibnamefont{Fleischmann}},
  \bibnamefont{and} \bibinfo{author}{\bibfnamefont{T.}~\bibnamefont{Geisel}},
  \bibinfo{journal}{Physical Review Letters} \textbf{\bibinfo{volume}{111}},
  \bibinfo{pages}{013901} (\bibinfo{year}{2013}).

\bibitem[{\citenamefont{Ying et~al.}(2011)\citenamefont{Ying, Zhuang, Heller,
  and Kaplan}}]{Ying:2011fp}
\bibinfo{author}{\bibfnamefont{L.~H.} \bibnamefont{Ying}},
  \bibinfo{author}{\bibfnamefont{Z.}~\bibnamefont{Zhuang}},
  \bibinfo{author}{\bibfnamefont{E.~J.} \bibnamefont{Heller}},
  \bibnamefont{and} \bibinfo{author}{\bibfnamefont{L.}~\bibnamefont{Kaplan}},
  \bibinfo{journal}{Nonlinearity} pp. \bibinfo{pages}{R67--R87}
  (\bibinfo{year}{2011}).

\bibitem[{\citenamefont{Nye}(1999)}]{nye1999natural}
\bibinfo{author}{\bibfnamefont{J.~F.} \bibnamefont{Nye}},
  \emph{\bibinfo{title}{{Natural Focusing and Fine Structure of Light: Caustics
  and Wave Dislocations}}} (\bibinfo{publisher}{Institute of Physics
  Publishing}, \bibinfo{year}{1999}).

\bibitem[{\citenamefont{Kulkarny and White}(1982)}]{Kulkarny:1982tm}
\bibinfo{author}{\bibfnamefont{V.~A.} \bibnamefont{Kulkarny}} \bibnamefont{and}
  \bibinfo{author}{\bibfnamefont{B.~S.} \bibnamefont{White}},
  \bibinfo{journal}{Physics of Fluids} \textbf{\bibinfo{volume}{25}},
  \bibinfo{pages}{1770} (\bibinfo{year}{1982}).

\bibitem[{\citenamefont{Metzger et~al.}(2010)\citenamefont{Metzger,
  Fleischmann, and Geisel}}]{metzger_universal_2010}
\bibinfo{author}{\bibfnamefont{J.~J.} \bibnamefont{Metzger}},
  \bibinfo{author}{\bibfnamefont{R.}~\bibnamefont{Fleischmann}},
  \bibnamefont{and} \bibinfo{author}{\bibfnamefont{T.}~\bibnamefont{Geisel}},
  \bibinfo{journal}{Physical Review Letters} \textbf{\bibinfo{volume}{105}},
  \bibinfo{pages}{020601} (\bibinfo{year}{2010}).

\bibitem[{\citenamefont{De~Haan and Ferreira}(2006)}]{DeHaan:2006bi}
\bibinfo{author}{\bibfnamefont{L.}~\bibnamefont{De~Haan}} \bibnamefont{and}
  \bibinfo{author}{\bibfnamefont{A.}~\bibnamefont{Ferreira}},
  \emph{\bibinfo{title}{{Extreme Value Theory: An Introduction}}}, Springer
  Series in Operations Research and Financial Engineering
  (\bibinfo{publisher}{Springer New York}, \bibinfo{year}{2006}).

\bibitem[{\citenamefont{Albeverio and Piterbarg}(2006)}]{Albeverio:2006up}
\bibinfo{author}{\bibfnamefont{S.}~\bibnamefont{Albeverio}} \bibnamefont{and}
  \bibinfo{author}{\bibfnamefont{V.}~\bibnamefont{Piterbarg}}, in
  \emph{\bibinfo{booktitle}{Extreme events in nature and society}}, edited by
  \bibinfo{editor}{\bibfnamefont{S.}~\bibnamefont{Albeverio}},
  \bibinfo{editor}{\bibfnamefont{V.}~\bibnamefont{Jentsch}}, \bibnamefont{and}
  \bibinfo{editor}{\bibfnamefont{H.}~\bibnamefont{Kantz}}
  (\bibinfo{publisher}{Springer}, \bibinfo{year}{2006}), pp.
  \bibinfo{pages}{47--68}.

\end{thebibliography}

\end{document}